\documentstyle[aps,prl,12pt,epsf]{revtex}

\begin{document}
\title{Ordering and Excitations in the Field-Induced Magnetic Phase
of Cs$_{3}$Cr$_{2}$Br$_{9}$}

\author{B. Grenier$^{1}$, Y. Inagaki$^{2}$, L.P. Regnault$^{1}$, A. Wildes$^{3}$, T. Asano$^{2}$,
Y. Ajiro$^{2}$, E. Lhotel$^{4}$, C. Paulsen$^{4}$, T. Ziman$^{3}$
and J.P. Boucher$^{5}$.}

\address{$^{1}$D\'{e}partement de Recherche Fondamentale
sur la Mati\`{e}re Condens\'{e}e, SPSMS, Laboratoire
Magn\'{e}tisme et Diffraction Neutronique, CEA-Grenoble, 38054
Grenoble cedex 9, France.\\ $^{2}$Department of Physics, Kyushu
University, Fukuoka 850-8581, Japan.\\ $^{3}$Institut Laue
Langevin, BP 156, 38042 Grenoble cedex 9, France.\\ $^{4}$Centre
de Recherches sur les Tr\`{e}s Basses Temp\'{e}ratures CNRS, BP
166, 38042 Grenoble cedex 9, France.\\ $^{5}$Laboratoire de
Spectrom\'{e}trie Physique, Universit\'{e} J. Fourier, BP 87,
38402 Saint Martin d'H\`{e}res, France.}

\date{June 2003}
\maketitle

\

\begin{abstract}
Field-induced magnetic order has been investigated in detail in
the interacting spin $3/2$ dimer system Cs$_{3}$Cr$_{2}$Br$_{9}$.
Elastic and inelastic neutron scattering measurements were
performed up to $H = 6$ T, well above the critical field
$H_{c1}\sim 1.5$ T. The ordering displays incommensurabilities and
a large hysteresis before a commensurate structure is reached.
This structure is fully determined. Surprisingly, the lowest
excitation branch never closes. Above $H_{c1}$, the gap increases
slowly with field. An analysis in terms of projected pseudo-spins
is given.
\end{abstract}

\


\

 \narrowtext

In recent years, a great deal of work has been devoted to
isotropic quantum spin systems which show a gap in their energy
spectrum. This includes a large variety of different models such
as Haldane and alternating chains, ladders, as well as spin-dimer
systems. In zero magnetic field, these systems are all
characterized by a $S=0$ singlet ground state ($|0\rangle$) and
the energy gap $E_{G}$ of the lowest magnetic excitation
corresponds to a $S=1$ state ($|1\rangle$). In a field $H$, when
the gap closes at $H=H_{c1}$ ($g\mu _{B}H_{c1}\sim E_{G}$) a
three-dimensional magnetic ordering develops, induced by the small
inter-chain and/or inter-dimer couplings. In these field-induced
magnetic orderings (FIMO) only the spin components transverse to
the applied field spontaneously break symmetry. Moreover, as
initially proposed for Haldane chains, a FIMO transition is an
experimental realization of a Bose-Einstein condensation (BEC) of
the hard-sphere bosons - the magnons - into the ground-state \cite
{Affleck}. Recently, the magnetic behavior observed in the
$s=\frac{1}{2}$ spin-dimer compound TlCuCl$_{3}$ has been analyzed
within this framework \cite {Nikuni,Ruegg}. While for BEC a
gapless Goldstone mode is present in the energy spectrum of the
condensed phase, in the case of a FIMO, the mode is gapless only
if there is conservation of the component of the total spin in the
field direction. If this symmetry is broken, the total number of
the magnons is no longer conserved. A gap may then be seen in the
lowest excitation branch. It is important to understand the way an
anisotropy creates such a gap in a FIMO phase. In this letter, we
present a complementary elastic and inelastic neutron
investigation of the FIMO phase in the spin-dimer system
Cs$_{3}$Cr$_{2}$Br$_{9}$. This material differs from TlCuCl$_{3}$
and similar compounds \cite{Kato} in two significant respects. The
larger spin value of Cr$^{3+}$, $s=3/2$ favors the presence of
local anisotropies. The crystal structure is different: the
spin-dimers form an hexagonal arrangement in the ({\bf a}, {\bf
b}) plane as shown in Fig. 1. This leads to frustration, which
should strongly affect any magnetic order. These two points, i.e.,
the role of a small anisotropy and the effects of frustration in a
FIMO are, here, demonstrated experimentally. An analysis in terms
of projected pseudo-spins is proposed to discuss the striking
features in the magnetic phase.

\begin{figure}[tbp]
\centerline{\epsfxsize=70mm \epsfbox{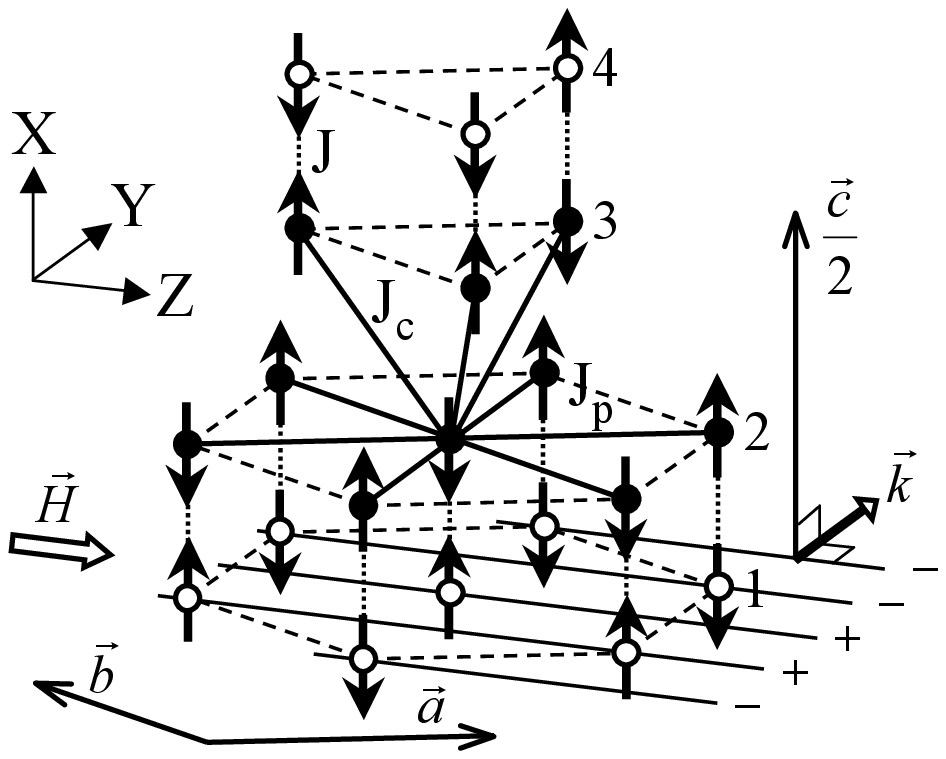}} \vspace{+0.3
cm}\caption{Transverse spin order of the Cr$^{3+}$ ions in the
commensurate $++--$ magnetic phase of Cs$_3$Cr$_2$Br$_9$ for ${\bf
H}\parallel {\bf a}-{\bf b}$. ${\bf k}$ is the propagation vector
of the magnetic order.}
\end{figure}

The spin-excitation spectrum of Cs$_{3}$Cr$_{2}$Br$_{9}$ in zero
field was determined across the whole Brillouin zone by
Leuenberger et al. \cite{Leuenberger1}. The intra- and inter-dimer
exchange couplings are all {\it antiferromagnetic} (AF). In a
field, Zeeman splitting was also observed, but only for small
fields, $H\leq H_{c1}$ \cite{Leuenberger2}. Neither the
field-induced magnetic structure nor the field dependence of the
spin excitations have been studied above $H_{c1}$. Here, we
investigate this FIMO phase in detail. Our experimental
configuration is the same as in Ref. \cite{Leuenberger2}: the
external field $H$ is applied within the ({\bf a}, {\bf b}) plane
along the {\bf a} - {\bf b} axis (hereafter defined as the ${\bf
Z}$ axis) so that the $XY$ plane where the AF ordered moments are
expected to lie ({\it transverse ordering}) is defined by the {\bf
c } (hereafter {\bf X}) and the {\bf a} + {\bf b} ({\bf Y}) axes
(Fig. 1). Cs$_{3}$Cr$_{2}$Br$_{9}$ crystallizes in the hexagonal
$P6_{3}/mmc$ structure, with $a=b=7.508$ \AA , $c=18.70$ \AA\
(which is twice the distance between successive planes), and
consists of four Bravais sublattices. The $s=3/2$ Cr$^{3+}$ dimers
lie parallel to the {\bf c} axis \cite{Structure}.

\begin{figure}[tbp]
\centerline{\epsfxsize=70mm \epsfbox{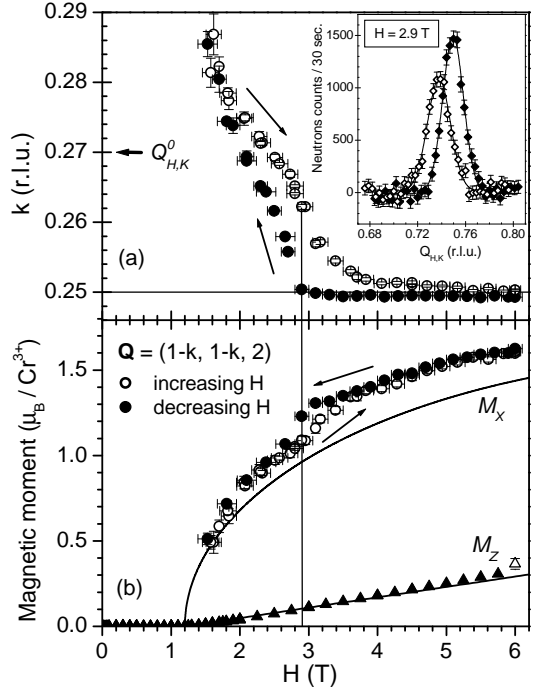}} \vspace{0.3 cm}
\caption{a) Field-dependence of the propagation wave-vector$\!$
component $k$ ($=k_{X}=k_{Y}$) and b) that of the parallel (${\cal
M}_{Z}$) and the ordered transverse (${\cal M}_{X}$)
magnetizations. $Q_{H,K}^{o}$ is the wave vector corresponding to
the minima of the dispersions (see Fig. 3 and text). In panel 2b,
the solid lines are theoretical predictions obtained for the
commensurate $++--$ magnetic order (see text).}
\end{figure}

First, we present the elastic results. They were performed on the
lifting-arm two-axis diffractometer D23 CEA-CRG at the Institut
Laue-Langevin (ILL) in Grenoble (France). A single crystal ($\sim
350$ mm$^{3}$) of Cs$_{3}$Cr$_{2}$Br$_{9}$ was placed in a 6T
vertical cryomagnet equipped with a dilution insert. The
wavelength was $\lambda =1.28 $ \AA\ and the temperature was
maintained in the range $50-100$ mK. In zero field, $113$ nuclear
reflections were collected, which could be reduced to $68$
independent measurements. The structural refinement was achieved
with type II Becker-Coppens Gaussian extinction correction and
isotropic thermal factors. A weighted R-factor of $1.7$ \% on the
nuclear structure factors F$_N$ was obtained by refining $12$
parameters. The magnetic structure was then determined at $H=6$ T.
It is described by the commensurate propagation wave vector ${\bf
k}={\bf (}1/4{\bf ,}$ $1/4{\bf ,}$ $0{\bf )}$ \cite{wave vector}.
Due to the symmetry, two other magnetic domains were observed,
corresponding to the wave vectors ${\bf (}1/4{\bf ,}$ $-1/2{\bf
,}$ $0{\bf )}$, ${\bf (}-1/2,$ $1/4,$ $0{\bf)}$. $123$ magnetic
reflections were collected in the three domains. For the magnetic
refinement an isotropic Cr$^{3+}$ magnetic form factor was used,
yielding a weighted R-factor of $4.2$ \% on the magnetic structure
factors F$_M$. This gives: {\it {\bf {i)}}} the populations of the
three domains are about the same ($34.8\pm 0.6$, $33.0\pm 0.4$ and
$32.2\pm 0.4\%)$ and the magnetic structure is collinear with the
moments pointing along the ${\bf c}$ ($\equiv {\bf X}$) direction;
this agrees with the expected transverse ordering for a FIMO, {\it
{\bf {\ ii)}}} the arrangement of the transverse magnetic moments
within one cell is represented in Fig. 1 (atoms labelled 1, 2, 3,
4). If we assume the moment amplitude to be constant (as is usual
for 3d ions), this motif propagates with the sequence $++--$ along
the $(110)$ direction (for the first domain), as shown in Fig. 1,
and after refinement, we obtain that the AF ordered transverse
moment is ${\cal M}_{X}=1.60(1)$ $\mu _{B}$ per Cr$^{3+}$ ion at
$H=6$ T. Considering the black atoms in this figure, {\it
ferromagnetic} correlations between neighboring spins, i.e., {\it
frustration}, are seen both within the ({\bf a},{\bf b}) plane and
between the planes. The field-dependence (for $H<6$ T) of the
propagation wave-vector component $k=k_{X}=k_{Y}$ and that of
${\cal M}_{X}$ are displayed in Fig. 2 (open and solid circles).
These data were from scans in the $(110)$ direction across the
strongest magnetic peak $(1-k,$ $1-k,$ $2)$ with $k\sim 0.25$, for
increasing and decreasing magnetic field $H$. When increasing
(decreasing) $H$, this peak could be detected from (down to)
$H\sim 1.5$ T: $H_{c1}$ is, then, slightly smaller than this
value. Each scan was well fitted by a Gaussian function, giving
rise to an accurate evaluation of both the integrated intensity,
i.e., ${\cal M}_{X}^{2}$, and the wave-vector component $k$ (see
the inset in Fig. 2a where the magnetic peak is recorded for the
same field value, $H=2.9$ T, while $H$ is increased or decreased).
In Cs$_{3}$Cr$_{2}$Br$_{9}$, within the ordered phase, there are
both incommensurabilities and large hysteresis effects suggesting
pinning \cite{McMillan}. For increasing $H$, the commensurate
$++--$ structure is reached at $H\sim 5$ T, while for decreasing
$H$, the commensurate structure remains locked down to $H\sim 3$
T. The {\it parallel} magnetization ${\cal M}_{Z}(H)$ was measured
both by neutrons from the ferromagnetic contribution appearing on
top of the weak nuclear Bragg peak ($1,$ $1,$ $4$) at $H=6$ T (the
open triangle in Fig. 2b) and from SQUID at $T\sim 100$ mK in the
full field range (solid triangles).

Second, we report inelastic results. The same crystal was mounted
on the cold neutron triple axis spectrometer IN12 FZ/CEA-CRG at
ILL, in a $6$T vertical cryomagnet with a dilution insert. The
measurements were performed in the same experimental configuration
at $T\sim 50$ mK. The final wave-vector ${\bf k}_{f}$ was fixed to
various values between $1.07$ and $1.55$ \AA $^{-1}$, depending on
the explored energy range and on the energy resolution that was
needed. Examples of the dispersion as a function of
$Q_{H,K}=Q_{H}=Q_{K}$ (at $Q_{L}=2$) are shown in Fig. 3.

\begin{figure}[tbp]
\centerline{\epsfxsize=70mm \epsfbox{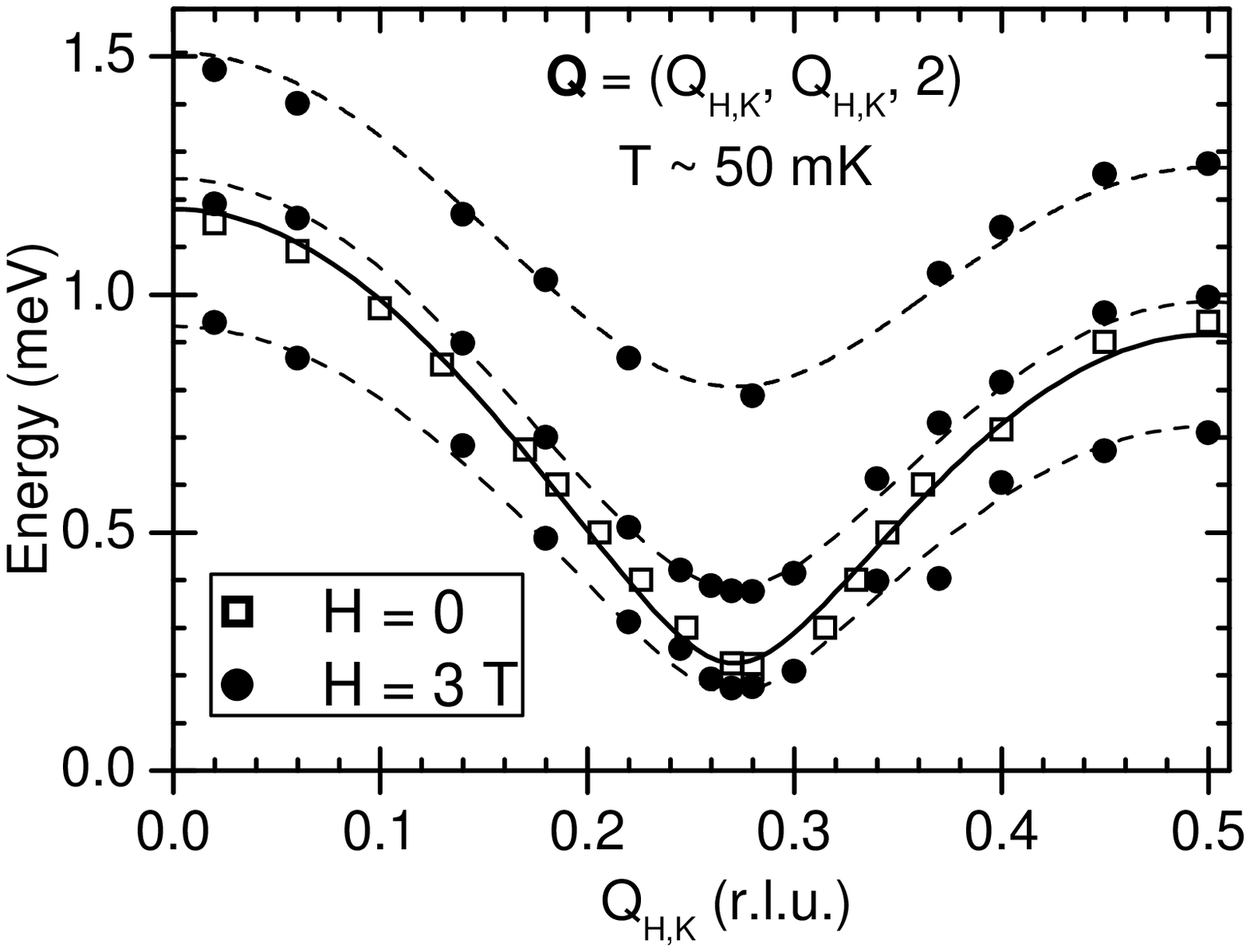}} \vspace{0.3 cm}
\caption{Examples of dispersions for $H = 0$ and $H = 3$ T. The
solid line for $H = 0$ is a fit as in Ref. [$5$]. The dashed lines
for $H = 3$ T are guides to the eye.}
\end{figure}

For $H=0$ (open squares), a unique excitation branch is observed,
while three distinct branches are seen in the ordered phase (solid
circles for $H=3$ T). The description proposed in Ref.
\cite{Leuenberger1} for $H=0$ reproduces very well the data (see
solid line). From this fit, we obtain the following exchange
values: $J=1.03$ meV for the intra-dimer coupling, $J_{p}=0.054$
and $J_{c}=0.039$ meV for the in-plane and out-of-plane
inter-dimer interactions, respectively, in agreement with Ref.
\cite{Leuenberger1}. We now focus on the minima of the
dispersions, i.e., the three ${\it energy}$ ${\it gaps}$. They
always occur at the same wave-vector value $Q_{H,K}^{o}\sim 0.27$
r.l.u., even above $H_{c1}$ where the propagation vector component
$k$ varies continuously from $0.29$ down to $0.25$ r.l.u. (see
Fig. 2a). The field-dependence of these gaps, obtained from energy
scans performed at ($0.27$, $0.27$, $2$), is plotted in Fig. 4.
The inset shows such a scan recorded at $H=0$ with the best energy
resolution ($k_{f}=1.07$ \AA $^{-1}$): a splitting - {\it not
detected previously} \cite{Leuenberger1} - is clearly observed
\cite {Comment1}. It reveals the presence of an additional
anisotropy which can be described by a single-ion term
$D(s_{i}^{X})^{2}$ (${\bf X}\equiv {\bf c}$) at each Cr$^{3+}$
site. Such a term results in a splitting of the initial
dispersion. This effect can be accounted for by redefining for
each branch two distinct effective intra-dimer couplings
\cite{Leuenberger3}: $J^{-}=J+0.8D$ and $J^{+}=J-1.6D$ (for
$s=3/2$ and $D < 0$). From the splitting at the minimum of the
dispersions, we evaluate $D\sim -0.01$ meV. With this value, the
Zeeman splitting below $H_{c1}$ (solid lines in Fig. 4) agrees
also very well with the data. From the lowest energy branch, the
critical field is expected at $H_{c1}\sim 1.5$ T, in agreement
with the diffraction result. As mentioned above, for $H>H_{c1}$,
three distinct branches are still observed. The two upper branches
agree roughly with ``Zeeman'' behaviors: $g\mu _{B}H$ and $2g\mu
_{B}H$ with $g=2$ (dotted lines). The evidence for a gap (${\cal
E}_{g}$) on the lowest energy branch above $H_{c1}$ is, however, a
new result. It shows that {\it there is no massless Goldstone mode
in the FIMO phase of} $Cs_{3}Cr_{2}Br_{9}$. In fact, even near
$H_{c1}$ the lowest energy gap does not vanish. The half-width
($0.05$ meV) of the fitted gaussian at the minimum gap $\sim 0.11$
meV is instrumental.

\begin{figure}[tbp]
\centerline{\epsfxsize=70mm \epsfbox{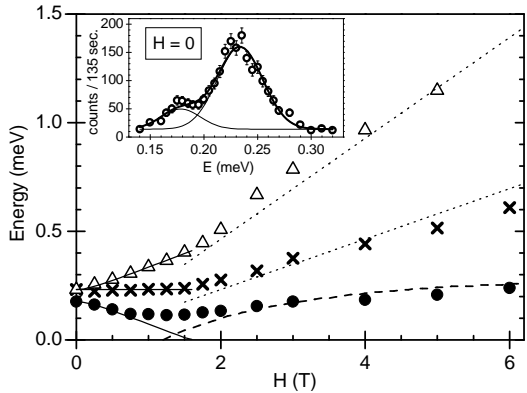}} \vspace{0.3 cm}
\caption{Field dependence of the energy gaps. The solid lines
describe the splitting below $H_{c1}$. Above $H_{c1}$, the dotted
lines are $g\mu_{B}H$ and $2g\mu_{B}H$. The dashed line describes
the lowest gap ${\cal E}_{g}$ (see text). Inset: H = 0 energy scan
performed at (0.27, 0.27, 2)with $k_{f}=1.07$ \AA $^{-1}$. The
splitting signals the presence of a small single-ion anisotropy
($D$ in Eq. 1). The slightly broader width of the upper mode
indicates the possible presence of small higher-order
anisotropies.}
\end{figure}

Including the single-ion anisotropy $D$, the Hamiltonian for the
spin-dimers in Cs$_{3}$Cr$_{2}$Br$_{9}$ is

\begin{eqnarray}
{\cal H} &=&\sum_{i}[J{\bf s}_{ia}.{\bf s}
_{ib}+D(s_{ia}^{X^{2}}+s_{ib}^{X^{2}})-g\mu
_{B}H(s_{ia}^{Z}+s_{ib}^{Z})] \nonumber \\ &&+\sum_{i,m}J^{\prime
}({\bf s}_{ia}.{\bf s}_{ma}+{\bf s}_{ib}.{\bf s}_{mb}) \eqnum{1}
\end{eqnarray}

\noindent with $J^{\prime }=J_{p}$ ($J_{c}$) for in (out-of)-plane
couplings. In Eq. (1), the indices $a$ and $b$ refer to the two
spins in each dimer while the index $m$ accounts for the
interacting spins surrounding each spin $i$ (6 in-plane and 2x3
out-of-plane couplings). Because of the spin value $s=3/2$, three
distinct plateaus are, a priori, expected in the magnetization
curve of such a compound. They are actually observed in the parent
compound Cs$_{3}$Cr$_{2}$Cl$_{9}$, and magnetic ordered phases are
expected to precede each of the three plateaus \cite{Inagaki}. The
situation is different in Cs$_{3}$Cr$_{2}$Br$_{9}$ where the
magnetization constantly increases from $H_{c1}\sim 1.3$ T up to
saturation ($H_{s}\sim 26$ T) without intermediate plateaux
\cite{Inagaki}. This means that the three FIMO phases in this
compound overlap giving rise to a single FIMO phase, hereafter
called the {\it extended }FIMO phase. In general, a {\it simple}
FIMO phase develops each time a crossing occurs between two states
of the isolated dimers. In low field, $H\sim H_{c1}$, the crossing
occurs between the singlet ground-state $|0\rangle$ and the lowest
Zeeman split state $|1,-1\rangle$ (including anisotropy) of the
initial triplet state $|1\rangle$. For a simple FIMO phase, an
effective Hamiltonian ${\cal H}_{eff}$ can be derived by
projecting Eq. (1) onto these elementary states. One obtains a
$\sigma =\frac{1}{2}$ pseudo-spin hamiltonian, which, for the
initial $s=3/2$ spin system, is given to first order:

\begin{eqnarray}
{\cal H}_{eff} &=&\sum_{i,m}J_{eff}^{XY}(\sigma _{i}^{X}\sigma
_{m}^{X}+\sigma _{i}^{Y}\sigma _{m}^{Y})+J_{eff}^{Y}\sigma
_{i}^{Z}\sigma _{m}^{Z}  \nonumber \\ &&+d_{eff}(\sigma
_{i}^{X}\sigma _{m}^{X}-\sigma _{i}^{Y}\sigma
_{m}^{Y})-\sum_{i}g\mu _{B}H_{eff}\sigma _{i}^{Z}  \eqnum{2}
\end{eqnarray}

\noindent with $J_{eff}^{XY}\!\!=\!\!2M^{2}J^{\prime }$,
$J_{eff}^{Z}\!\!=\!\!J^{\prime }(c_{1}^{2}\!-\!c_{-1}^{2})^{2}/2$,
$d_{eff}\!\!=$ $-4M^{2}J^{\prime }c_{1}c_{-1}$ and $g\mu
_{B}H_{eff}\!=\!g\mu _{B}H(c_{1}^{2}-c_{-1}^{2})+J$$+J^{\prime
}(c_{1}^{2}\!\!-\!\!c_{-1}^{2})/4\!\!-\!\!R^{2}D.$ In these
expressions, $ M\!\!=$ $^{o}\langle 1,-1|s_{X}|0\rangle
\!=\!-\sqrt{5/2}$, $R\!=\!-2(1-6c_{1}c_{-1})/5$, $c_{1}$ and
$c_{-1}$ account for the mixing of the initial states (no
anisotropy):
$|1,-1\rangle=c_{-1}|1,-1\rangle^{o}+c_{1}|1,1\rangle^{o}$ with
$c_{-1}=\beta D/\left[ (\beta D)^{2}+\{[(g\mu _{B}H)^{2}+(\beta
D)^{2}]^{1/2}-g\mu _{B}H\}^{2}\right] ^{1/2}$ and $c_{1}=\{[(g\mu
_{B}H)^{2}+(\beta D)^{2}]^{1/2}-g\mu _{B}H\}/$ $\left[ (\beta
D)^{2}+\{[(g\mu _{B}H)^{2}+(\beta D)^{2}]^{1/2}-g\mu
_{B}H\}^{2}\right] ^{1/2}$ where $\beta
=2\,\,^{o}\langle1,-1|s_{X}^{2}|1,+1\rangle^{o}=6/5$. We note that
the ratio $J_{eff}^{Z}/J_{eff}^{XY}\sim 1/(4M^{2})=1/10$ is much
smaller than in the $s=1/2$ spin case \cite{Tachiki}, i.e., for
$s=3/2$, the pseudo-spin hamiltonian is almost XY. More novel is
the XY anisotropy term $d_{eff}$. This anisotropy is induced by
$D$, but it occurs indirectly through the mixing coefficient
$c_{1}c_{-1}$. Note that higher-order anisotropies within the YZ
hexagonal plane (see Fig. 4 caption) would renormalize slightly
the ratio $J_{eff}^{Z}/J_{eff}^{XY}$ and $d_{eff}$, but otherwise
does not affect the analysis. According to Eq. (2), in a {\it
simple} FIMO phase, one has an anisotropic quantum pseudo-spin
$1/2$ XYZ model. For an {\it extended} FIMO phase - as in
Cs$_{3}$Cr$_{2}$Br$_{9}$ - the pseudo-spin description should give
the essential physics, but one expects deviations. We now compare
our data and predictions from Eq. (2). For the $\sigma $ spin
system, we analyze the commensurate $++--$ structure of the FIMO
phase. The effective couplings in the mean-field approximation
according to the spin arrangement displayed in Fig. 1 (see the
black atoms) are $\langle J_{eff}^{XY}\rangle
=-4(J_{p}+J_{c}/2)M^{2}$ and $\langle d_{eff}\rangle =$
$+4(J_{p}+J_{c}/2)M^{2}c_{1}c_{-1}$. The calculated ${\cal M}_{Z}$
and ${\cal M}_{X}$, with ${\cal M}_{X}=2|M|\sqrt{{\cal
M}_{Z}(1-{\cal M}_{Z})}$, are represented by the solid lines in
Fig. 2b. The observed behaviors are well explained qualitatively.
Quantitatively, at $H=6$ T, the discrepancy between measurements
and this simple theory is $\sim 15$ \%\ for ${\cal M}_{X}$. Such a
correction is what one expects when one compares the results for a
simple and an extended FIMO, i.e., when the contributions from the
higher states are taken into account \cite{Tim}. Extrapolation of
these curves to lower fields predicts a reasonable critical value,
$H_{c1}\sim 1.2$ T.

A FIMO corresponds to an {\it AF spin-flop} state. A small
anisotropy, such as $d_{eff}$ in Eq. (2), breaks the axial
symmetry responsible for the Goldstone mode and opens a gap in the
XY fluctuations. By Holstein-Primakoff transformation on the
pseudo-spin hamiltonian, extending the standard treatment
\cite{SFAF} to include $d_{eff}$:

\begin{eqnarray}
{\cal E}_{g} &\sim &4\;\sigma _{X}^{2}\sqrt{\langle
J_{eff}^{XY}\rangle \langle d_{eff}\rangle} \nonumber
\\
&=&16\;\sigma _{X}^{2}(J_{p}+J_{c}/2)M^{2}\sqrt {\left|
c_{1}c_{-1}\right|} \eqnum{3}
\end{eqnarray}

\noindent where $\sigma _{X}={\cal M}_{X}/(2\left| M\right|) $ is
the ordered transverse component of the $\sigma =1/2$ spins
($0<\sigma _{x}<\sigma=1/2)$. The prediction from Eq. (3) applies
to a simple FIMO. In the case of an extended FIMO, one expects
additional contributions resulting from the mixing of states. An
``amplification'' factor is therefore to be expected. As shown by
the dashed line in Fig. 4, with a multiplying factor $\sim 5$, Eq.
(3) provides a quite good agreement with the data. This factor is
surprisingly large and suggests that the dynamics is much more
sensitive to the extended nature of the FIMO than the statics.

At $H=H_{c1}$, a FIMO transition is expected to be of second
order. The transition here is complicated by the frustration: the
magnons that soften are incommensurate and degenerate in momentum
space \cite{Leuenberger1}. The initial ordering vector appears
away in wave vector space from the minimum gap, which, remarkably,
remains non-zero at the transition. The complexity seen in our
results as to the evolution of the ordering vector towards a
locked-in commensurate value with collinear spins, is, we believe,
due to interplay of quantum fluctuations, enhanced by the
frustration of the hexagonal lattice and the anisotropy. Even in
the commensurate phase a small anisotropy, since it opens a gap,
changes the nature of the magnon condensation.

We have observed new and peculiar features at the onset of
field-induced magnetic order in Cs$_{3}$Cr$_{2}$Br$_{9}$. They
should stimulate new developments, concerning simple versus
extended FIMO, incommensurability and hysteretic behavior.

We thank E. Ressouche and J. Rodriguez Carvajal for help in the
structural refinement.


\begin{references}
\bibitem{wave vector} Throughout the text, the wave vectors are
expressed in reciprocal lattice units (r.l.u.).
\bibitem{Comment1}  In Fig. 3, the data for $H=0$ were obtained with
$k_{f}=1.25$ \AA $^{-1}$: the splitting of each mode could not be
seen.
\end{references}
\end{document}